\documentclass[12pt,preprint]{aastex}
\slugcomment{}

\begin{document}

\title{Stochastic Gravitational Wave Background from Neutron Star r-mode Instability Revisited}

\author{Xing-Jiang Zhu$^{1}$, Xi-Long Fan$^{2,3}$ and Zong-Hong Zhu$^{1}$}

\affil{$^1$ Department of Astronomy, Beijing Normal University,
Beijing 100875, China; zhuzh@bnu.edu.cn \\
$^2$ Dipartimento di Fisica, Sezione di Astronomia, Universit\`a di
Trieste, via G.B. Tiepolo 11, I-34131, Trieste, Italy \\
$^3$ I.N.A.F. Osservatorio Astronomico di Trieste, via G.B. Tiepolo
11, I-34131, Trieste, Italy}

\begin{abstract}
We revisit the possibility and detectability of a stochastic
gravitational wave (GW) background produced by a cosmological
population of newborn neutron stars (NSs) with r-mode instabilities.
The NS formation rate is derived from both observational and
simulated cosmic star formation rates (CSFRs). We show that the
resultant GW background is insensitive to the choice of CSFR models,
but depends strongly on the evolving behavior of CSFR at low
redshifts. Nonlinear effects such as differential rotation,
suggested to be an unavoidable feature which greatly influences the
saturation amplitude of r-mode, are considered to account for GW
emission from individual sources. Our results show that the
dimensionless energy density $\Omega_{\rm{GW}}$ could have a peak
amplitude of $\simeq (1-3.5) \times10^{-8}$ in the frequency range
$(200-1000)$~Hz, if the smallest amount of differential rotation
corresponding to a saturation amplitude of order unity is assumed.
However, such a high mode amplitude is unrealistic as it is known
that the maximum value is much smaller and at most $10^{-2}$. A
realistic estimate of $\Omega_{\rm{GW}}$ should be at least 4 orders
of magnitude lower ($\sim 10^{-12}$), which leads to a pessimistic
outlook for the detection of r-mode background. We consider
different pairs of terrestrial interferometers (IFOs) and compare
two approaches to combine multiple IFOs in order to evaluate the
detectability of this GW background. Constraints on the total
emitted GW energy associated with this mechanism to produce a
detectable stochastic background (a SNR of 2.56 with 3-year cross
correlation) are $\sim 10^{-3}\hspace{1mm} M_{\odot} c^2$ for two
co-located advanced LIGO detectors, and $2 \times 10^{-5}
\hspace{1mm} M_{\odot} c^2$ for two Einstein Telescopes. These
constraints may also be applicable to alternative GW emission
mechanisms related to oscillations or instabilities in NSs depending
on the frequency band where most GWs are emitted.
\end{abstract}

\keywords{gravitational waves - stars: neutron - stars: formation -
supernovae: general}

\section{Introduction}
\label{intro} A stochastic gravitational wave background (SGWB) is a
target for gravitational wave (GW) interferometers (IFOs). It could
have two very different origins. It may result from a large variety
of cosmological processes developed in the very early universe, such
as amplification of quantum vacuum fluctuations, phase transitions,
cosmic strings, etc. (see, e.g., Maggiore 2000; Buonanno 2003 for
reviews). Additionally, an astrophysical GW background (AGWB) is
expected to be produced by the superposition of a large number of
unresolved sources since the beginning of star formation (Schneider
et al. 2000; Regimbau \& Mandic 2008). There have been a host of
literatures dedicated to the studies of various AGWB sources, such
as core collapse supernovae (CCSNe) \cite{blair}, leading to the
formation of neutron stars (NSs) (Coward et al. 2001; Howell et al.
2004; Buonanno et al. 2005) or black holes (BHs) (Ferrari et al.
1999a; de Araujo et al. 2004; Pereira \& Miranda 2009), phase
transitions in NSs (Sigl 2006; de Araujo \& Marranghello 2009),
coalescing compact binaries consisting of NSs and/or BHs (Schneider
et al. 2001; Farmer \& Phinney 2002; Regimbau \& de Freitas Pacheco
2006a; Regimbau \& Chauvineau 2007), magnetars (Regimbau \& de
Freitas Pacheco 2006b) and population III stars (Sandick et al.
2006; Suwa et al. 2007; Marassi et al. 2009) among others.

In this paper we revisit the possibility that the r-mode
instabilities in newly born NSs could form a SGWB. NSs, having long
been considered to be likely observational sources for GW detection,
emit gravitational radiation in a number of ways, for example,
through CCSNe, inspiralling compact binaries, rotating deformed
stars, oscillations and instabilities \cite{GW_NS}. First postulated
more than ten years ago (Andersson 1998; Friedman \& Morsink 1998),
the r-mode instability has been attracting increased attention due
to the fact that it is driven unstable by GW emission and it can be
active for a wide range of core temperatures and angular velocities
(Lindblom et al. 1998; Andersson \& Kokkotas 2001; Andersson et al.
2003). Early estimate indicated an energy equivalent to roughly
$1\%$ of a solar mass is radiated in GWs (Andersson et al. 1999) as
an initially rapidly rotating star spins down. This led to an
expectation of a SGWB produced by a cosmological population of young
rapidly rotating NSs with closure density $h^2\Omega_{\rm{GW}}$
peaking at $\sim 10^{-8}$ of the present-day critical energy density
of the universe \cite{Owen1998, Ferrari1999b}.

The most important aspect of the r-mode intability is the largest
amplitude (often called the saturation amplitude $\alpha$) that the
perturbation can grow to. This maximum amplitude determines how fast
the NS spins down and whether the associated GW emission will be
detectable (either in terms of single event or a stochastic
background). In Owen et al. (1998) and then Ferrari et al. (1999b)
it was taken to be of order unity (there were no estimates of this
maximum at that time). Later more detailed (both analytical and
numerical) studies seriously questioned the potential of the
instability as well as the efficiency of GW emission (Rezzolla et
al. 2000; Ho \& Lai 2000; Lindblom et al. 2000; Rezzolla et al.
2001a, 2001b; Lindblom \& Owen 2002). On the one hand, it was
suggested that energy transfer to other stellar inertial modes can
significantly reduce the saturation amplitude of r-mode
\cite{Schenk2002,Morsink2002,Brink2004}. Arras et al. (2003) tested
nonlinear coupling between stellar inertial modes and revealed much
lower values of saturation amplitude ($\alpha \sim 10^{-4} -
10^{-1}$). Then a specific resonant three-mode coupling between the
r-mode and the pair of fluid modes was identified to be responsible
for the catastrophic decay of large-amplitude r-modes and a
perturbative analysis of the decay rate suggested a maximum
dimensionless saturation amplitude $\alpha_{\rm{max}} < 10^{-3} -
10^{-2}$ \cite{Lin2006}. More recently Bondarescu et al. (2009)
examined the 3-mode coupling between the r-mode and two other
inertial modes and showed that the r-mode evolution can progress in
a number of different directions depending on unknown properties of
the viscosity, leading to very complex consequent mode evolution and
the associated GW signal.

On the other hand, differential rotation, first suggested by
Rezzolla et al. (2000, 2001a, 2001b), is an unavoidable feature of
nonlinear r-modes (Stergioulas \& Font 2001; Lindblom et al. 2001;
S{\'a} 2004). Small values of $\alpha$ mentioned above are also
supported by studies on the role of differential rotation, causing
large scale drift of fluid elements, in the nonlinear evolution of
r-modes (S\'a \& Tom\'e 2005). In particular they parametrize the
initial amount of differential rotation by $K$ and then relate the
parameter $K$ to the largest amplitude that the r-mode can grow to.
In this paper we will use the characteristic GW amplitude given by
S\'a \& Tom\'e (2006) to account for the average source spectrum.
The adopted GW amplitude, parametrized by parameter $K$, in fact
scales with the saturation amplitude $\alpha$. Below we will discuss
the influence of this quantity on the r-mode background.

In addition to the average source spectra, the properties of AGWB
also depend on GW source formation rate. Studies of cosmic star
formation rate (CSFR) allow estimation of the birth rate of NSs. In
the last decade our knowledge of cosmic star formation has been
greatly improved due to advances in astronomical observation and
hydrodynamic simulation. Here we take into account both
observational and simulation-based CSFR models to obtain the NS
formation rate and discuss their effects on the resultant GW
background. In particular, we will investigate the role of the
maximal redshift of different CSFR models in our results.

The high-frequency window of GW spectrum ($10~\rm{Hz}\leq f \leq$ a
few kHz) is open today through pioneering efforts of the
first-generation terrestrial IFOs, such as Laser Interferometer
Gravitational Wave Observatory (LIGO) (Abramovici et al. 1992) in
Livingston (LIGOL) and in Hanford (LIGOH), Virgo (Caron et al. 1997)
near Pisa, GEO600 (L\"uck et al. 1997) in Hanover and TAMA300 (Ando
et al. 2001) at Tokyo. Although GWs were not detected, an
observational upper limit ($\Omega_{\rm{GW}} <6.9 \times 10^{-6}$)
was placed on the energy density of SGWB at around 100~Hz, exceeding
previous indirect limits from the Big Bang Nucleosynthesis and the
cosmic microwave background \cite{LIGO limit}. In the future the
SGWB from NS r-mode instability, among others, may offer an
important detection target for the proposed second and third
generation detectors represented by advanced
LIGO\footnote{http://www.ligo.caltech.edu/advLIGO/} (or advanced
Virgo\footnote{http://wwwcascina.virgo.infn.it/advirgo/}) and the
Einstein Telescope (ET\footnote{http://www.et-gw.eu/}) respectively.

Detectors throughout the world can act as a network in order to
improve the detection ability to the SGWB. Two approaches of
combining 2N IFOs are proposed in Allen \& Romano (1999): (i)
correlating the outputs of a pair of IFOs, then combining the
multiple pairs, and (ii) directly combining the outputs of 2N IFOs.
For any given real\footnote{Here by ``real" we mean considering the
real overlap reduction functions for different detector pairs other
than assuming an optimal value of unity since this function plays a
crucial role in determining the frequency-dependent sensitivity of
each detector pair to the stochastic background (Finn et al. 2009).
Results in the present paper also show that using an optimal value
of unity can lead to overestimates in signal-to-noise ratios of more
than one order of magnitude.} IFOs it is necessary to compare these
two optimal approaches of detecting the SGWB. Cella et al. (2007)
has shown that the approach of combining multiple pairs of IFOs
using Virgo, LIGO and GEO can improve the detection ability to the
SGWB by simulating an isotropic GW background with an
astrophysically-motivated spectral shape. Fan \& Zhu (2008) compared
the detection ability of the two approaches for stochastic GWs from
string cosmology.

Detectability of the r-mode background are demonstrated here by
calculating signal-to-noise ratios (SNRs) for pairs of currently
operating IFOs and advanced detectors at their design sensitivities.
We also consider two approaches of combining 4 real IFOs to examine
how many improvements can be obtained. The organization of this
paper is as follows. In Section 2 we review works on the
determination of CSFR and present five CSFR models. In Section 3 we
derive the NS formation rate as a function of redshift using the
adopted CSFR models. Then by combining the source formation rate
from Section 3 and the characteristic GW amplitude of individual
events, spectral properties of the r-mode stochastic background are
investigated in Section 4. We will discuss the detectability of
r-mode background in Section 5 and finally Section 6 is devoted to
our conclusions.

Throughout the paper, the so-called 737 $\Lambda$CDM cosmology is
assumed with $H_{0}=100 h\cdot \rm{km}\hspace{0.5mm}
\rm{s}^{-1}\hspace{0.5mm} \rm{Mpc}^{-1}$ with $h=0.7$ and
$\Omega_{m}=0.3$, $\Omega_{\Lambda}=0.7$ (e.g., Komatsu et al.
2009).

\section{Cosmic star formation rate}
The CSFR, which has tight connection with GW event rate, is of
intense interest to many fields of astrophysics. For many years
effort has gone into studying the  cosmic star formation history
(see, e.g., Madau et al. 1996; Hopkins 2004; Wilkins et al. 2008).
Since CSFR is not a directly observable quantity, usually the
rest-frame ultraviolet (UV) light is considered to be an indicator
(see Calzetti 2008 and references therein for details about CSFR
indicators) of star formation because it is mainly radiated by
short-lived massive stars. With the help of Hubble Space Telescope
(HST) and other large telescopes, galaxy luminosity density of
rest-frame UV radiation is studied, and then converted into CSFR
density through the adoption of a universal stellar initial mass
function (IMF) to calculate the conversion factor. Many authors have
developed parameterized fits to the expected evolution of the CSFR
with redshift. Firstly, following Porciani \& Madau (2001), we adopt
three different forms which model the CSFR density for redshifts up
to $z \approx 4$:

\begin{equation}
\dot{\rho}_{\ast}(z)_{i}=1.67 C_{i}h_{65}F(z) G_{i}(z) \quad
M_{\odot} \hspace{0.5mm}\rm{yr}^{-1} \hspace{0.5mm}\rm{Mpc}^{-3},
\end{equation}
with i=1, 2, 3 denoting the different models, $C_i$ a constant,
$h_{65}=h/0.65$, $G_{i}(z)$ a function of z and
$F(z)=[\Omega_{m}(1+z)^3 + \Omega_{\Lambda}]^{1/2}/(1+z)^{3/2}$. A
constant factor of $1.67$ is applied to account for conversion of a
Salpeter IMF with a lower cutoff from $0.5 M_{\odot}$ to $0.1
M_{\odot}$ (the one we will use below). $F(z)$ convert the assumed
cosmology from an Einstein-de Sitter universe to the $\Lambda$CDM
cosmology (Porciani \& Madau 2001). The first fit (hereafter SFR1)
is given by Madau \& Pozzetti (2000), with $C_{1}=0.3$ and
$G_{1}(z)= e^{3.4z} /(e^{3.8z}+45)$ where the CSFR increases rapidly
from $z=0$ to reach a peak at around $z=1.5$ and then gradually
declines at higher redshifts. The second one (SFR2) is from Steidel
et al. (1999) with $C_{2}=0.15$ and $G_{2}(z)= e^{3.4z}
/(e^{3.4z}+22)$ where the CSFR remains roughly constant at $z\geq2$.
The third model (SFR3) from Blain et al. (1999) has $C_{3}=0.2$ and
$G_{3}(z)= e^{3.05z-0.4} /(e^{2.93z}+15)$ where CSFR increases at
higher redshifts to account for effects of dust extinction.

With the improvement in measurements of galaxy luminosity functions
at a broad range of wavelengths, star formation history can be
traced to higher redshifts. Here we consider the work by Hopkins \&
Beacom (2006), who refined the previous models up to redshift $z
\sim 6$ from new measurements of the galaxy luminosity function in
the UV (SDSS, GALEX, COMBO17) and far-infrared (FIR) wavelengths
(Spitzer Space Telescope). A parametric fit (hereafter HB06) is
given by:
\begin{equation}
\dot{\rho}_{\ast}(z)=h\frac{0.017+0.13z}{1+(z/3.3)^{5.3}}\quad
M_{\odot} \hspace{0.5mm}\rm{yr}^{-1} \hspace{0.5mm}\rm{Mpc}^{-3},
\end{equation}
assuming 737 cosmology and a modified Salpeter A IMF \cite{IMF}.
Although the IMF used to derive HB06 is different from the standard
Salpeter's, this will not introduce considerable errors to our
results because the evolution of CCSNe rate based on the CSFR and on
an assumed universal IMF is largely independent of the choice of the
IMF (Madau 1998).

Many authors have addressed the issues of calibrating
 the high-z
CSFR through long-duration gamma-ray bursts (GRBs) (see, e.g.,
Y\"uksel et al. 2008 and Kistler et al. 2009). Recently Wang $\&$
Dai (2009) uses latest GRBs data to constrain the CSFR up to
$z=8.3$. Meanwhile there are other methods to determine the high-z
CSFR, such as observations of color-selected Lyman break galaxies
(LBGs) (Bouwens et al. 2008) and $Ly\alpha$ emitters (Ota et al.
2008). However, such calibrations cannot reach considerable
agreements except for an overall decline at $z\geq4$ (see Figure 1
of Y\"uksel et al. 2008, Kistler et al. 2009 and Wang \& Dai 2009).
Due to huge uncertainties and the incompleteness of data sets we
will not include them here.

On the other hand, Springel \& Hernquist (2003) derive the CSFR from
hydrodynamic simulation of structure formation in $\Lambda$CDM
cosmology. They study the history of cosmic star formation from the
``dark ages", at redshift $z=20$ to the present. The CSFR obtained
in their study is broadly consistent with measurements given
observational uncertainty and can be remarkably well-fitted by the
following form (hereafter SH03):
\begin{equation}
\dot{\rho}_{\ast}(z)=\dot{\rho}_{m}\frac{\beta
\exp[\alpha(z-z_{m})]}{\beta - \alpha + \alpha \cdot
\exp[\beta(z-z_{m})]},
\end{equation}
where $\alpha=3/5$, $\beta=14/15$, $z_{m}=5.4$ marks a break
redshift and $\dot{\rho}_{m}=0.15$ $M_{\odot}
\hspace{0.5mm}\rm{yr}^{-1} \hspace{0.5mm}\rm{Mpc}^{-3}$ fixes the
overall normalization. It is worth mentioning that they consider a
$\Lambda$CDM model with the same parameters with our assumed 737
cosmology.

In Fig. 1 we plot the CSFR predicted in the above five models as a
function of redshift. SFR1, SFR2 and SFR3 show distinguishable
features at $z\geq2$. SH03 peaks at a much higher redshift, between
$z=5$ and $z=6$, than observation-based models (around $z\simeq2$).
The cutoff of each curve in Fig. 1 corresponds to maximum redshifts
of CSFR models: $z_{\star}=4$ for SFR1-3, $z_{\star}=6$ for HB06 and
$z_{\star}=20$ for SH03. What Fig. 1 illustrates is our poor
understanding about star formation history at high redshifts from
astronomical observations.

Note that there are some other CSFR models, similar to or different
from the five models adopt here, not included since our aim is not
to make a complete survey on this issue but to phenomenologically
investigate its influences on the SGWB from an ensemble of
astrophysical sources. We refer readers to Calura \& Matteucci
(2003), Daigne et al. (2004), Bromm \& Loeb (2006), Nagamine et al.
(2006) and Fardal et al. (2007) for details of other studies on the
determination of CSFR. In the following sections we will investigate
how different CSFRs affect the rate of NS formation and spectral
properties of AGWB.

\section{Neutron star formation rate}
Since the evolving rate of CCSNe closely tracks the star formation
rate, using CSFR models presented in Section 2 we can estimate the
number of NSs formed per unit time within the comoving volume out to
redshift z (Ferrari et al. 1999a):
\begin{equation}
R_{NS}(z)=\int_{0}^{z}\dot{\rho}_{\ast}(z') \frac{dV}{dz'} dz'
\int_{m_{\rm{min}}}^{m_{\rm{max}}}\Phi(m)dm \label{dR_NS},
\end{equation}
where $\dot{\rho}_{\ast}(z)$ is the CSFR density, $dV/dz$ is the
comoving volume element, and $\Phi(m)$ is the IMF. Here we assume
that each CCSN results in either a NS or a BH and take a NS
progenitor mass range of $8 M_{\odot} - 25 M_{\odot}$. In order to
make comparison with Ferrari et al. (1999b) we also consider a lower
upper limit for NS progenitor masses $m_{\rm{max}}=20 M_{\odot}$ as
indicated from core collapse simulations by Fryer (1999). However
according to Belczynski \& Taam (2008) the mass of NS progenitor
might be greater than $40 M_{\odot}$ for stars in a binary system.
So we will also include a higher limit of $m_{\rm{max}}=40
M_{\odot}$ for our calculations of NS formation rate (In Sigl 2006
the progenitor mass to form a NS ranges from $10 M_{\odot}$ to $40
M_{\odot}$).

Note that in some studies (e.g., Coward et al. 2001; de Araujo et
al. 2004; Regimbau \& Mandic 2008) with respect to AGWB there is an
additional $(1+z)$ term in Eq.(\ref{dR_NS}) dividing the CSFR to
account for the time dilatation of CSFR due to cosmic expansion.
Here we do not include such a term according to de Araujo \& Miranda
(2005) who argue that the inclusion of this additional term is
inadequate.

To integrate through Eq.(\ref{dR_NS}) one still needs to know the
forms of $dV/dz$ and $\Phi(m)$. Following Regimbau \& Mandic (2008),
the comoving volume element is related to z through
\begin{equation}
\frac{dV}{dz}=4\pi\frac{c}{H_{0}}\frac{r(z)^{2}}{E(\Omega ,z)},
\end{equation}
where $H_{0}$ the Hubble constant, $E(\Omega
,z)=\sqrt{\Omega_{\Lambda}+\Omega_{m}(1+z)^{3}}$ and $r(z)$ the
comoving distance related to the luminosity distance by
$d_{L}=r_{z}(1+z)$.

We consider the standard Salpeter IMF: $\Phi(m)=Am^{-(1+x)}$ with
$x=1.35$, where A is a normalization constant, obtained through the
relation $\int_{m_{1}}^{m_{u}}m\Phi(m)dm=1$ with $m_{1}=0.1
M_{\odot}$ and $m_{u}=125 M_{\odot}$. Then we plot the NS formation
rate $R_{\rm{NS}}(z)$ defined in Eq.(\ref{dR_NS}) for the CSFR
models presented in Section 2 with a modest value of
$m_{\rm{max}}=25 M_{\odot}$ in Fig. 2. Note that SFR1, SFR2 and SFR3
show quite different behaviors for $z>2.5$ and observation-based
models give rise to more NS formation than SH03 up to respective
redshift limits, but SH03 predicts a much higher cumulative NS
formation rate for $z\geq 10$.

In Table 1 we present the total number (per unit time) of CCSN
explosions leaving behind a NS out to corresponding redshift limits
for the five CSFR models, and for three values of $m_{\rm{max}}$ :
$20 M_{\odot}$, $25 M_{\odot}$ and $40 M_{\odot}$. We compare the
results obtained here with those in Ferrari et al. (1999b), and find
a factor of $\sim2-3$ enhancement for the total NS formation rate,
which is mainly due to differences of CSFR models and cosmology
terms (e.g., different forms for the comoving volume element).
\begin{table}
\caption {Total rate of NS formation in five CSFR models for three
values of the upper limits of NS progenitor masses.}
\begin{center}
\begin{tabular}{lccc}
\hline
 Model (redshift limit) & $m_{\rm{max}}=20 M_{\odot}$ & $25 M_{\odot}$ & $40 M_{\odot}$ \\
\hline
 SFR1 \, ($z_{\star}=4$) & 30.0 & 33.2 & 37.4 \\
 SFR2 \, ($z_{\star}=4$) & 39.3 & 44.5 & 49.1 \\
 SFR3 \, ($z_{\star}=4$) & 47.2 & 52.2 & 58.9 \\
 HB06 \, ($z_{\star}=6$) & 47.5 & 52.6 & 59.3 \\
 SH03 \, ($z_{\star}=20$) & 62.0 & 68.6 & 77.4 \\
\hline
\end{tabular}
\end{center}
\end{table}

\section{Spectral properties of the SGWB from NS r-mode instability}
In this section we will evaluate the spectral properties of the
stochastic background produced by an ensemble of newly born NSs with
nonlinear r-mode instabilities. Initially, let us review the
formalism used to characterize the AGWB.

It is useful to characterize the spectral properties of a SGWB by
specifying how the energy is distributed in frequency domain.
Explicitly, one introduces a dimensionless quantity,
$\Omega_{\rm{GW}}$ given by:
\begin{equation}
\Omega_{\rm{GW}}(\nu_{\rm{obs}})=\frac{1}{\rho_{c}}\frac{d\rho_{\rm{GW}}}{d\ln\nu_{\rm{obs}}},
\end{equation}
where $\rho_{\rm{GW}}$ is the GW energy density, $\nu_{\rm{obs}}$
the frequency in the observer frame and $\rho_{c}=3H_{0}^{2}/8\pi G$
is the critical energy density required to close the universe today.
For a stochastic background of astrophysical origin, the energy
density is given by:
\begin{equation}
\Omega_{\rm{GW}}(\nu_{\rm{obs}})=\frac{\nu_{\rm{obs}}}{c^{3}\rho_{c}}F_{\nu}(\nu_{\rm{obs}}),
\end{equation}
where the spectral density of the flux at the observed frequency
$\nu_{obs}$ is defined as
\begin{equation}
F_{\nu}=\int f_{\nu}(\nu_{\rm{obs}})dR,
\end{equation}
where $f_{\nu}(\nu_{\rm{obs}})$ is the energy flux per unit
frequency (in $\rm{erg}\cdot \rm{cm}^{-2}\cdot \rm{Hz}^{-1}$)
produced by a single source and $dR$ is the differential GW event
rate.

The energy flux per unit frequency $f_{\nu}(\nu_{\rm{obs}})$ can be
written as follows (Carr 1980)
\begin{equation}
f_{\nu}(\nu_{\rm{obs}})=\frac{\pi c^{3}}{2G}h_{c}^{2} \label{flux},
\end{equation}
where $h_{c}$ is the dimensionless amplitude produced by an event
that generates a signal with observed frequency $\nu_{\rm{obs}}$.

In order to obtain the spectral properties (e.g., the values of
$\Omega_{\rm{GW}}$ as a function of $\nu_{\rm{obs}}$) of the r-mode
stochastic background, we have the differential rate of source
formation $dR_{\rm{NS}}(z)$ through Eq.(\ref{dR_NS}) and still need
to know the energy flux emitted by a single source.

It has been shown that differential rotation can significantly
influence the detectability of GWs emitted by a spinning-down
newborn NS due to r-mode instability (S\'a \& Tom\'e 2006). Studies
of mode-mode coupling in rotating stars also indicate that the
maximum amplitude that r-mode can grow to is much smaller than
previously estimated (Arras et al. 2003). Here we use the
characteristic GW amplitude given by S\'a \& Tom\'e (2006):
\begin{equation}
h_{c}(\nu)=\frac{5.5\times10^{-22}}{\sqrt{K+2}}\sqrt{\frac{\nu}{\nu_{\rm{max}}}}\left(
\frac{20Mpc}{d_{L}}\right) \label{amplitude},
\end{equation}
where $K$ is a constant giving the initial amount of differential
rotation associated with the r-mode, and lies in the interval
$-5/4\leq K \leq 10^{13}$ (see S\'a \& Tom\'e (2006) for details),
$\nu=\nu_{\rm{obs}}(1+z)$ is the frequency in the source frame,
$\nu_{\rm{max}}$ the maximum frequency of emitted GWs given by
$2\Omega_{\rm{K}}/3\pi$ where $\Omega_{\rm{K}}=5612$ Hz is the
Keplerian frequency at which the star starts shedding mass at the
equator, assumed to be the initial value of the angular velocity of
the star, and $d_{L}$ is the luminosity distance to the source. Note
that the saturation amplitude $\alpha \propto (K+2)^{-1/2}$ (S\'a \&
Tom\'e 2006), which means the GW amplitude is proportional to
$\alpha$, same as those given in Bondarescu et al. (2009). In the
following calculations we should keep in mind that the parameter $K$
used in this paper is equivalent to the saturation amplitude
$\alpha$ of NS r-mode instability.

From the above equations we can obtain the dimensionless energy
density:

\begin{equation}
\Omega_{\rm{GW}}(\nu_{\rm{obs}})= {4\pi^{2}(1.1\times 10^{-20})^{2}
\over
3H_{0}^{2}(K+2)}\frac{\nu_{\rm{obs}}^{2}}{\nu_{\rm{max}}}\times
\bigg[\int_{z_{\rm{min}}}^{z_{\rm{max}}}
\int_{m_{\rm{min}}}^{m_{\rm{max}}} \dot{\rho}_{\ast}(z) (1+z)
\bigg(\frac{1 \rm{Mpc}}{d_{L}}\bigg)^{2} {dV\over dz}
\Phi(m)dmdz\bigg] \label{Omgw}.
\end{equation}

Thus, by setting a value for $K$ one can calculate $\Omega_{GW}$
numerically through Eq.(\ref{Omgw}) combined with corresponding
equations for CSFR, comoving volume element and IMF. Here we set
$m_{\rm{min}}=8 M_{\odot}$ and $m_{\rm{max}}=25 M_{\odot}$, while
$z_{\rm{min}}$ and $z_{\rm{max}}$ can be determined in such a way:
since frequencies of emitted GWs in the source frame range from
$\nu_{\rm{min}}=77-80$ Hz to
$\nu_{\rm{max}}=2\Omega_{\rm{K}}/3\pi=1191$ Hz, where the minimum
frequency corresponds to the final angular velocity of the star -
$0.065 \Omega_{\rm{K}}$ for $K =-5/4$ and $0.067 \Omega_{\rm{K}}$ if
$K \gg 1$, we have $\nu_{\rm{min}}/(1+z)\leq \nu_{\rm{obs}}\leq
\nu_{\rm{max}}/(1+z)$, which means sources with different redshifts
that produce a signal at the same frequency $\nu_{\rm{obs}}$ should
meet the condition: $\nu_{\rm{min}}/\nu_{\rm{obs}}-1\leq z \leq
\nu_{\rm{max}}/\nu_{\rm{obs}}-1$. Besides, we consider signals
emitted at early epochs up to the present ($z\geq 0$) and take into
account the maximal redshift ($z_{\ast}$) of CSFR model. Then we
obtain $z_{\rm{min}}= \rm{max}(0,\nu_{\rm{min}}/\nu_{\rm{obs}}-1)$,
$z_{\rm{max}}= \rm{min}(z_{\ast},\nu_{\rm{max}}/\nu_{\rm{obs}}-1)$,
which is similar to that of Owen et al. (1998) where $z_{\ast}\simeq
4$ is considered to be the maximum redshift where there was
significant star formation.

In Fig. 3 we plot the dimensionless energy density
$\Omega_{\rm{GW}}$ calculated for the five CSFR models presented in
Section 2 by setting $K$ at its minimal value: $K=-5/4$
corresponding to the smallest amount of differential rotation at the
time when the r-mode instability becomes active. However, as
emphasized by S\'a \& Tom\'e (2005), if $K$ is small, namely,
$K\approx 0$, it is necessary to consider other nonlinear effects
like mode-mode couplings in the calculation of $\alpha$, which will
again limit the maximum r-mode amplitude to values much smaller than
unity \cite{Arras2003}. In this respect, our choice of a minimum $K$
results in an unrealistically high upper limit for r-mode
background.

It is worth noting from Fig. 3 that no obvious differences are
recorded for the three curves of SFR1, SFR2 and SFR3, and
observation-based CSFR models give rise to stochastic backgrounds
about two times stronger than that of SH03 over a broad frequency
band, although SH03 leads to a much higher NS formation rate. The
sharp contrast between Fig. 3 and Fig. 2 indicates that the main
contribution to the GW background comes from low-redshift sources
because those events happened at higher redshifts have minor
influences due to the inverse squared luminosity distance dependence
of the single event energy flux. For the same reason, poor
observational understanding of high-z star formation history (see
Fig. 1) is not severe to studies about AGWB here.

To assess the role of $z_{\ast}$ in our results, we choose SFR2
(since it remains constant for $z\geq 2$), set three values for
$z_{\ast}$ ($z_{\ast}=4, 10$ and $20$), and then plot the
$\Omega_{\rm{GW}}$ in Fig. 4. It is surprising that the three curves
exhibit almost the same pattern in the frequency range
$\nu_{\rm{obs}}\geq 100$ Hz, and extending the redshift limit from 4
to 20 results in a growth of the lower-frequency background.
Increasing $z_{\ast}$ will enhance the background at lower
frequencies and even enable some formerly ``unavailable"
low-frequency signals to emerge. This low-frequency GW ``tail" can
be accounted for by the contribution from high-redshift sources.
However if CSFR is much lower at high-z, this effect will be
negligible. Thus Fig. 4 further support the conclusions from Fig. 3
and indicate that the most significant contribution to an AGWB comes
from GW events occurring at redshifts $z\leq 4$.

From Eq. (\ref{Omgw}) we find that $\Omega_{\rm{GW}}$ depends on the
values of $\nu_{\rm{max}} \sim \Omega_{\rm{K}}$ and $K$
($\Omega_{\rm{GW}} \propto \frac{1}{K+2} \sim \alpha^2$). As
suggested by Ferrari et al. (1999b), the Keplerian velocity
$\Omega_{\rm{K}}$ may have a broad distribution due to different
masses and radii of rotating NSs. Here we arbitrarily set
$\nu_{\rm{max}}$ ranging from $1000$ Hz to $2000$ Hz. For $K $ we
set $ -5/4, 100$ and $10^4$ corresponding to $\alpha =1, 0.1$ and
$10^{-2}$ respectively. Then we adopt HB06 as the CSFR model (below
we will use only HB06) and plot $\Omega_{\rm{GW}}$ as a function of
observed frequency for different values of $\alpha$ and
$\nu_{\rm{max}}$ in Fig. 5. We can see a higher peak for
$\Omega_{\rm{GW}}$ when we increase the maximum emitting frequency,
while the r-mode background for smaller $\nu_{\rm{max}}$ is slightly
enhanced at lower frequencies ($\leq 400$ Hz). On the other hand,
increasing the amount of differential rotation significantly reduce
the closure density and then affect the detectability of r-mode
background as we will discuss later. Considering a maximum value of
$10^{-2}$ for $\alpha$, a realistic estimate of r-mode background
should have a energy density at most $\sim 10^{-12}$ like the lowest
curve shown in Fig. 5.

Another important quantity of the AGWB is the so-called duty cycle,
which classifies the stochastic backgrounds in terms of continuous
background, popcorn noise and short noise (Coward \& Regimbau 2006):
\begin{equation}
D=\int_{0}^{\infty} \bar{\tau}(1+z)dR_{\rm{NS}}(z),
\end{equation}
where $\bar{\tau}$ is the average time duration of the GW emission
from a single source at the source frame, which dilated to
$\bar{\tau}(1+z)$ by the cosmic expansion, and $dR_{\rm{NS}}(z)$ is
the differential rate of NS formation in Eq.(\ref{dR_NS}). It has
been suggested that differential rotation can remarkably influence
the long-term spin and thermal evolution of NSs by prolonging the
duration of the r-modes (Yu et al. 2009). In view of the prolonged
r-mode, the spinning-down phase can last even longer than 1 yr,
which indicates a duty cycle $>\sim 10^9$. In the next section, we
will discuss the detectability of this continuous GW background.

\section{Detectability}
\subsection{Detecting the r-mode background with a network of IFOs}
We cannot reach sufficient sensitivity for detection of a SGWB with
a single terrestrial IFO since the output of a detector is dominated
by the noise rather than by the signal due to the stochastic
background itself (Allen 1996a; Maggiore 2000). The optimal strategy
to search for a SGWB is to cross correlate measurements of two or
more detectors. It has been shown that after correlating signals of
two detectors for a (i) isotropic, (ii) unpolarized, (iii)
stationary, and (iv) Gaussian stochastic background (see Allen \&
Romano 1999 for discussions of these assumptions), the optimal SNR
during an integration time T (here we assume $T = 3
\rm{yr}\simeq10^8$ s) is given by an integral over frequency $f$:
\begin{equation}
\left( {S \over N} \right)^2 = {9 H_0^4 \over 50 \pi^4} T
\int_0^\infty df \> {\gamma^2 (f) \Omega_{\rm{GW}}^2(f) \over f^6
P_1(f) P_2(f)}\ \label{SNR},
\end{equation}
where $P_1(f)$ and $P_2(f)$ are the power spectral noise densities
of the two detectors and $\gamma(f)$ is the so-called overlap
reduction function, first calculated by Flanagan (1993). This is a
dimensionless function of frequency and determined by the relative
locations and orientations of two detectors. For $\gamma(f)$, we
refer readers to Flanagan (1993), Allen \& Romano (1999), Maggiore
(2000) for more details.

To assess the detectability of the r-mode background, we will
calculate the SNRs for several pairs of detectors for
$\Omega_{\rm{GW}}$ computed with HB06 CSFR model, $K =-5/4$ and
$\nu_{\rm{max}} =1191$ Hz (unless otherwise stated we use these
parameters in Section 5). Here we consider the four IFOs which are
in routine operations - LIGOH (4km), LIGOL (4km), Virgo (3km) and
GEO (600m), as well as the second generation detectors - advanced
LIGO and advanced Virgo. Design sensitivity curves of these
detectors are shown in Fig. 6. It is worth mentioning that the
first-generation GW interferometric detectors have taken data at, or
close to, their design sensitivities (Fairhurst et al 2009). In
particular, the design sensitivity curve for initial LIGO was almost
attained by its S5 run. In the following calculations we will use
real $\gamma(f)$ for different pairs\footnote{Data of locations and
orientations of are taken from Allen (1996b).} of IFOs unless
otherwise stated.

Two approaches of combining 2N detectors to improve the detection
ability to the SGWB are proposed in Allen \& Romano (1999): (i)
correlating the outputs of a pair of detectors, then combining
multiple pairs (combining pairs, ``c-p"), and (ii) directly
combining the outputs of 2N detectors (directly combining, ``d-c").
For the first approach, the squared SNR is given by:
\begin{equation}
\left( {S \over N} \right)^2_{optI} = \sum_{pair}\left( {S \over N}
\right)^2_{pair},
\end{equation}
and for the second one:
\begin{equation}
\left( {S \over N} \right)^2_{op\,tII}\approx {}^{(12)} \left( {S
\over N} \right)^2\ {}^{(34)} \left( {S \over N} \right)^2
\cdots{}^{(2N-1,2N)} \left( {S \over N} \right)^2 +\hbox{{\rm\ all
possible permutations}}\ .
\end{equation}

\begin{table}
\caption {Concerning the detection of the r-mode background, we
present the SNRs calculated for multiple detector pairs of LIGOH
(H), LIGOL (L), Virgo (V) and GEO (G) and two approaches of
combining the four IFOs - combining pairs (c-p) and directly
combining (d-c). Case 1 corresponds to first-generation detectors
and Case 2 adopts sensitivities of advanced LIGO and advanced Virgo,
while G refers to an assumed interferometer with the same
sensitivity as advanced Virgo at the GEO site.}
\begin{center}
\begin{tabular}{lcccc}
\hline
 Case & H-L & L-V & H-V & V-G \\
\hline
 1 & $1.8\times 10^{-3}$ & $1.0\times 10^{-3}$ & $8.5\times 10^{-4}$ & $2.6\times 10^{-4}$ \\
 2 & 0.40 & 0.21 & 0.17 & 0.25 \\
\hline
 Case & L-G & H-G & c-p & d-c \\
\hline
 1 & $1.4\times 10^{-4}$ & $9.7\times 10^{-5}$ & $2.3\times 10^{-3}$ & $5.0\times 10^{-7}$ \\
 2 & 0.16 & 0.11 & 0.58 & 0.11 \\
\hline
\end{tabular}
\end{center}
\end{table}

We show in Table 2 the SNRs calculated for different pairs of LIGOH,
LIGOL, Virgo and GEO, and for two combinations of these four IFOs.
We consider here two cases representing two real networks of
first/second generation IFOs: Case 1 is for these four IFOs with
design sensitivities; Case 2 consists of two advanced LIGO detectors
and two advanced Virgo detectors both with proposed sensitivities.
Note that SNRs in Table 2 are lower than unity even for pairs of
advanced detectors. The most promising one ($SNR = 0.58$) comes from
combining pairs of four advanced IFOs. If we assume an optimized
value of unity for $\gamma(f)$, which is only possible for
co-located GW detectors \cite{gamma1}, the SNR is $11.0$ and $4.1$
for a pair of advanced LIGO and advanced Virgo IFOs respectively. We
note that by considering new detectors with comparable sensitivities
to advanced LIGO, such as LCGT in Japan (Kuroda et al. 1999) and
AIGO in Australia (Blair et al. 2008), it could be possible to reach
a higher, but still not significant SNR with a network of
second-generation IFOs.

In order to obtain some detectable parameter space we need at least
one order of magnitude higher SNRs than those in Case 2 of Table 2.
Then we reduce the noise power spectral densities of advanced LIGO
and advanced Virgo by a factor of 10 by hand, and investigate the
role of differential rotation ($K$) and maximum emitting frequency
($\nu_{\rm{max}}$) in the detectability of the r-mode background. We
are motivated here by the fact that third-generation detectors like
ET can reach a sensitivity roughly an order of magnitude better than
that of advanced LIGO (Hild et al. 2008).

In Fig. 7 we plot the SNR as a function of $K$ for H-L, H-V and L-V
pairs. As a natural result from Fig. 5, the detectability of SGWB
from r-mode instability is drastically reduced to 0 as $K$
approaching 10. The higher SNR of H-L pair reflects the lower noise
level of advanced LIGO. Due to similarity of the overlap reduction
functions (see Fig. 2 of Fan \& Zhu 2008) no significant difference
is shown between L-V and H-V pairs.

Fig. 8 shows the SNR evolution with $\nu_{\rm{max}}$ for H-L, H-V
and L-V pairs. In contrast to Fig. 5, in which a higher peak value
of $\Omega_{\rm{GW}}$ was obtained for larger $\nu_{\rm{max}}$,
there are no identical features for SNR evolutions here. For H-L
pair, we note that SNR varies inversely as the increase of
$\nu_{\rm{max}}$. This can be explained that the low-frequency GW
background is enhanced for smaller $\nu_{\rm{max}}$ while the growth
of high-frequency background due to larger $\nu_{\rm{max}}$ is
suppressed by the $1/f^6$ term in Eq. (\ref{SNR}). On the other
hand, we can reach higher SNRs for H-V and L-V pairs for larger
$\nu_{\rm{max}}$ ($\geq 1600$ Hz). This unique feature can be
attributed to particular evolving behaviors of $\gamma(f)$ for the
two pairs (see again Fig. 2 of Fan \& Zhu 2008 and we will find that
$\gamma(f)$ is extremely close to zero at high frequencies for H-L,
while it still fluctuates above and below zero up to $1000$ Hz for
H-V and L-V pairs). This is strongly supported by three curves in
Fig. 9, where we plot the SNR as a function of $\nu_{\rm{max}}$ by
assuming $\gamma(f)=1$ , show exactly the same evolving pattern as
the curve of H-L in Fig. 8.

For a detection rate $90\%$ and a false alarm rate $10\%$, the total
optimal SNR threshold should be 2.56. In order to evaluate the
promise of detecting the r-mode background we present in Fig. 10 the
regions in the ($\nu_{\rm{max}}, K$) plane where SNR could be higher
than 2.56 for multiple detector pairs. It is shown that the
detectable parameter space is quite limited even for a real network
of third-generation IFOs. In particular, a strong constraint for ($K
< 0$ or equivalently $\alpha \sim 1$) is obtained. Meanwhile we find
that two approaches of combining multiple IFOs can effectively
improve \footnote{In fact such an improvement is negligible
considering that SNR is extremely sensitive to the saturation
amplitude of r-mode ($\propto \alpha^2$).} the detection ability as
compared to H-L pair, and ``c-p" method performs better than ``d-c"
approach in terms of detecting the r-mode background.

\subsection{The detection prospects of future detectors}
In this section we discuss further the detection prospects of a SGWB
from NS r-mode instability by networks of third generation
instruments.

Fig. 10 indicates that only when the initial amount of differential
rotation is near its minimum value ($K \simeq 0$), the r-mode
background could be detectable by a real network of third-generation
GW detectors. However, as emphasized by S\'a \& Tom\'e (2005), if
$K$ is near its minimum value, other nonlinear effects like
mode-mode couplings should be included in the calculation of
saturation amplitude $\alpha$. Still in this case, $\alpha$ will be
limited to values much smaller than unity \cite{Arras2003}. In this
respect the physically reasonable values of parameter $K$ introduced
in S\'a \& Tom\'e (2005) should be larger than $ 10^4 $ in order to
be consistent with a maximum saturation amplitude $\alpha \leq
10^{-2}$ \cite{Lin2006}.

It seems most probable that the r-mode background is not going to be
detectable. Given the relation $SNR \sim \Omega_{\rm{GW}} \sim
\alpha^{2}$, a $10^{-2}$ saturation amplitude (compared with a value
of order unity) will reduce the SNR by 4 orders of magnitude for any
measurements of SGWB from NS r-mode instability. In particular, to
reach the same SNR requires 10000 times improvements in
sensitivities of two detectors.

If we leave the small $\alpha$ issue (the dominant aspect) aside, it
is still quite pessimistic to detect the r-mode background as seen
from Table 2 and Fig. 10. This is due to the following facts: 1) the
closure density of this background peaks at much higher frequencies
than the most sensitive frequency band of ground-based detectors; 2)
the minimum frequency of r-mode GW signal $\nu_{\rm{min}} = 80$ Hz
just misses the frequency band (1-60 Hz) where the overlap reduction
functions of real detector pairs are significant; 3) we should not
forget that we assumed all NSs are born with angular velocities near
their maximal value $\Omega_{\rm{K}}$. This is not necessarily
always the case since it would make more sense to consider some
fraction of NSs are born with rapid spins (Owen et al. 1998).
Actually it has been suggested that most NSs are born with very
small rotation rates \cite{NS_spin}. Current population synthesis
studies favor spin periods of NSs at birth in the range from tens to
hundreds of milliseconds \cite{ott,Perna2008}.

Realistic overlap reduction functions are adopted here for multiple
ground-based IFOs. Recall that two co-located advanced LIGO
detectors ($\gamma(f)=1$, SNR=11) perform even better than combining
four IFOs with 10-fold better sensitivities (SNR=5.8). With this in
mind we also evaluate the detectability of r-mode background with
ET, assuming two detectors located in Cascina, of triangular shape
($60^{\circ}$ between the two arms) and separated by an angle of
$120^{\circ}$ \cite{Eric2010}. The $\gamma(f)$ benefits a lot from
this configuration, nearly linearly decreasing from $-0.372$ at 1000
Hz to $-0.375$ at 1 Hz. We adopt the ET-B sensitivity from Hild et
al. (2008). For 3-year integration, a SNR of 2.56 requires $K \leq
\sim 150$ corresponding to $\alpha \sim 0.1$.

Then we convert the constrains on $K$ or $\alpha$ to absolute
numbers: the total emitted GW energy associated with NS r-mode
instability that enable the stochastic background to be detectable
by future detectors. The energy flux of single event can also be
written as:
\begin{equation}
f(\nu_{\rm{obs}},z)=\frac{1}{4 \pi d_{L}(z)^{2}}
\frac{dE_{\rm{GW}}}{d\nu}(1+z) \label{flux2},
\end{equation}
where $dE_{\rm{GW}}/{d\nu}$ is the gravitational spectral energy.
Combined with Eq.(\ref{flux}) and Eq.(\ref{amplitude}) we can obtain
the spectral energy density and thus the total GW energy emitted by
individual sources that contribute to this background. We give the
results (required energy level in order to obtain a SNR of 2.56 by
3-year integration) as follows (in $M_{\odot} c^2$): i) $1.8 \times
10^{-3}$ for a real network of ``third-generation" IFOs; ii) $9.4
\times 10^{-4}$ for two co-located advanced LIGO detectors; iii) $2
\times 10^{-5}$ for two detectors with ET-B sensitivity.

\section{Conclusions}
We revisit the possibility and detectability of a SGWB produced by a
cosmological population of young NSs with an r-mode instabilities.
Source formation rate is accounted for by using a set of CSFR
models, both observational and simulated. Our results show that the
resultant GW background is insensitive to the choice of CSFR models
(although they predict quite different NS formation rates), but
dependent on the evolving behavior of CSFR at low redshifts ($z\leq
2$). This is in good agreement with that in Howell et al. (2004).
But here we further investigate the effect of the maximal redshift
of CSFR models on the AGWB and find that high-redshift ($z > 4$)
sources could form a GW ``tail" in the low-frequency side ($\leq 30$
Hz, this number depends on particular source spectrum and thus only
applies to r-mode background) and has no effect on the
high-frequency background. Such an effect will be negligible if
high-z CSFR is much smaller. Our Fig. 3 and Fig. 4 indicate that the
most significant contribution to the GW background of astrophysical
origin comes from GW events occurring at redshifts $z\leq 4$.

The characteristic GW amplitude parametrized with the initial amount
of differential rotation ($K$) during r-mode evolution is adopted as
the average GW signal for individual sources. While a minimum $K$
corresponding to a saturation mode amplitude $\alpha \sim 1$ is
assumed, the energy density $\Omega_{\rm{GW}}$ has a maximum
amplitude at around $3\times 10^{-8}$, agreeing well with those of
Owen et al. (1998) and Ferrari et al. (1999b). However since we know
that the maximum amplitude $\alpha$ that r-mode can grow to is at
most $10^{-3} - 10^{-2}$. This means the physically reasonable
values of parameter $K$ is at least $10^4$. Consequently a realistic
estimate of $\Omega_{\rm{GW}}$ for r-mode background should be at
most $\sim 10^{-12}$.

We further consider multiple IFOs, including the first-generation
ones that are in routine operations and upgraded counterparts, for
the detection of r-mode background. We also illustrate how a network
of ground-based IFOs could improve the detection ability to this GW
background. Since the detectability is dominated by the square of
saturation amplitude ($SNR \propto \alpha^2$), it is likely that the
r-mode background will not be detectable for any future detectors.
Still we give the constraints on the total emitted GW energy
associated with this mechanism to enable a detection with $90 \%$
detection rate ($SNR=2.56$) by 3-year cross correlation as: $\sim
10^{-3} \hspace{1mm} M_{\odot} c^2$ for two co-located advanced LIGO
detectors and $2 \times 10^{-5}$ for two IFOs with ET sensitivity.
Considering the relatively certain NS formation rate, these
constraints might be applicable to alternative emission mechanisms
associated with NS oscillations and instabilities. This requires
further investigation. The requirement on GW energy level could be
lower if more signals are emitted near the most sensitive frequency
band of ground-based detectors ($\sim 40-200$ Hz). Through
reasonable assumptions of average source spectra, the lowest
detectable (in terms of stochastic background) GW energy for
individual source will be $10^{-7} \hspace{1mm} M_{\odot} c^2$ for
third generation detectors like ET \cite{SN_limit}. Overall, for the
detection of SGWB from NS instabilities, more efficient emitters are
required (see, e.g., Andersson et al. 2010 for reviews of GW
emission from NSs and Kastaun et al. 2010 for details of a possible
more efficient mechanism - f-mode).

While the SGWB from NS r-mode instability is difficult to detect,
the associated GW signal is still detectable in terms of single
events \cite{owen}, although this possibility also depends strongly
on the saturation amplitude. For instance, initially it was believed
that GWs from r-mode instability in a newborn NS could be detected
by advanced LIGO out to a distance of 20 Mpc \cite{Owen2002}.
However even for the most optimistic case in Bondarescu et al.
(2009), the detectable distance for advanced LIGO is only 1 Mpc. The
LIGO Scientific Collaboration and Virgo Collaboration have already
performed many searches for periodic GWs from rapidly rotating NSs
including the first search targeting the youngest known NS -
Cassiopeia A \cite{ns}. Although no GWs have been detected, direct
upper limits on GW emission from known pulsars like the Crab pulsar
have beaten down the indirect spin-down limits \cite{crab}.

\section*{Acknowledgments}
This work was supported by the National Natural Science Foundation
of China under the Distinguished Young Scholar Grant 10825313 and
Grant 11073005, and by the Ministry of Science and Technology
national basic science Program (Project 973) under Grant No.
2007CB815401. We thank the anonymous referee for valuable comments
and useful suggestions which improved this work very much. ZXJ is
grateful to Yun Chen, Eric Howell, David Blair and Luciano Rezzolla
for helpful discussions, and to Tania Regimbau for providing data of
ET sensitivity and $\gamma$ function.

\begin{figure}
\plotone{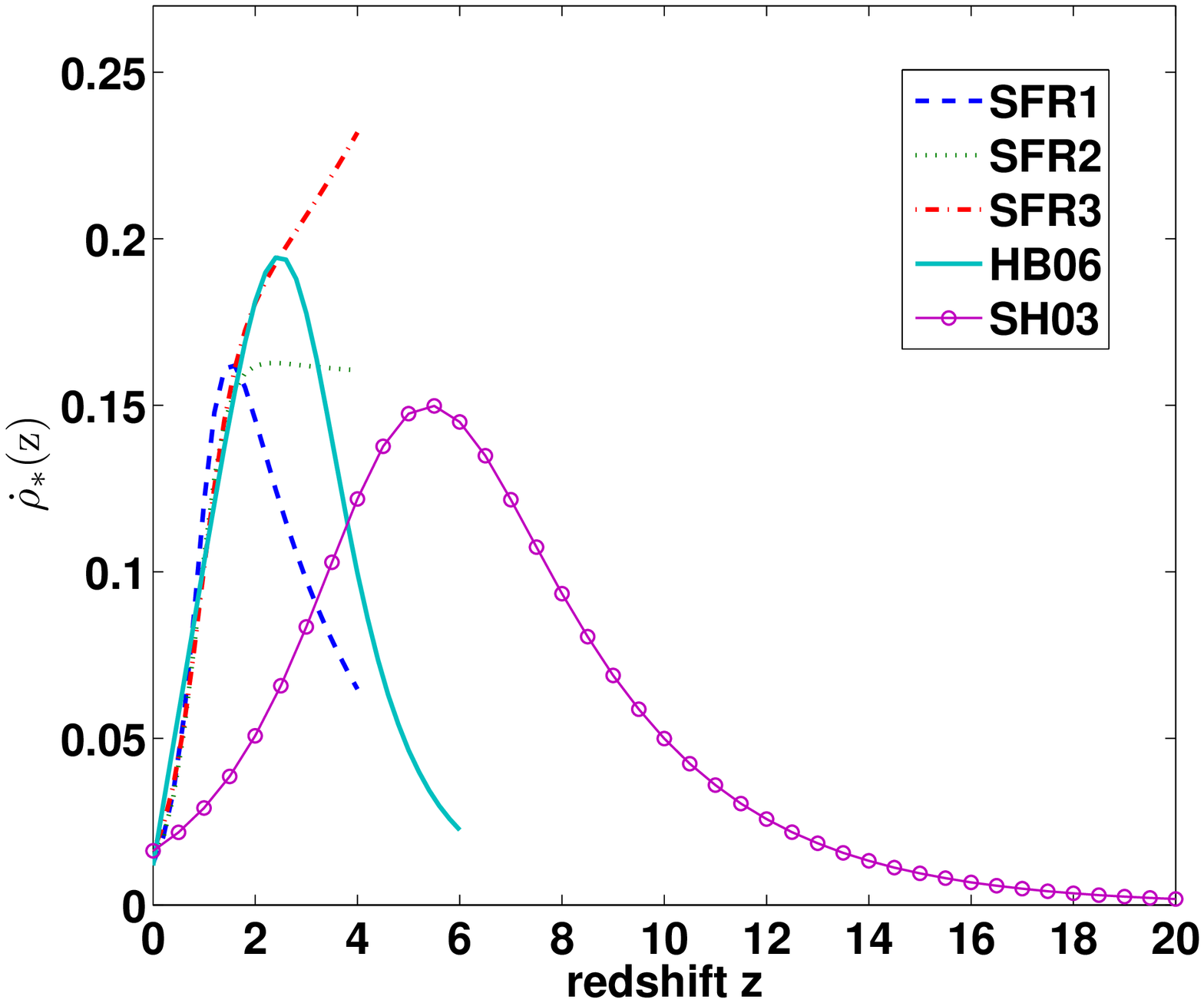} \figcaption{Evolution of the CSFR density
$\dot{\rho}_{\ast}(z)$ (in $M_{\odot} \hspace{0.5mm}\rm{yr}^{-1}
\hspace{0.5mm}\rm{Mpc}^{-3}$) predicted in five parameterized models
(see text).}
\end{figure}

\begin{figure}
\plotone{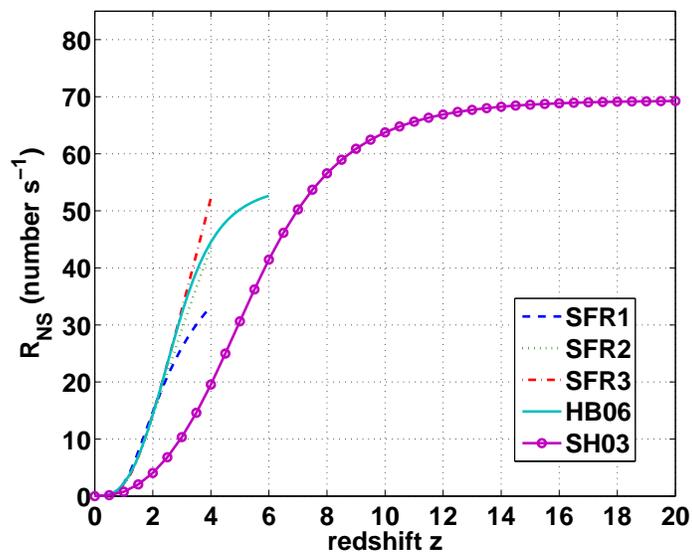} \figcaption{Evolution of the NS formation rate
under different CSFR models.}
\end{figure}

\begin{figure}
\plotone{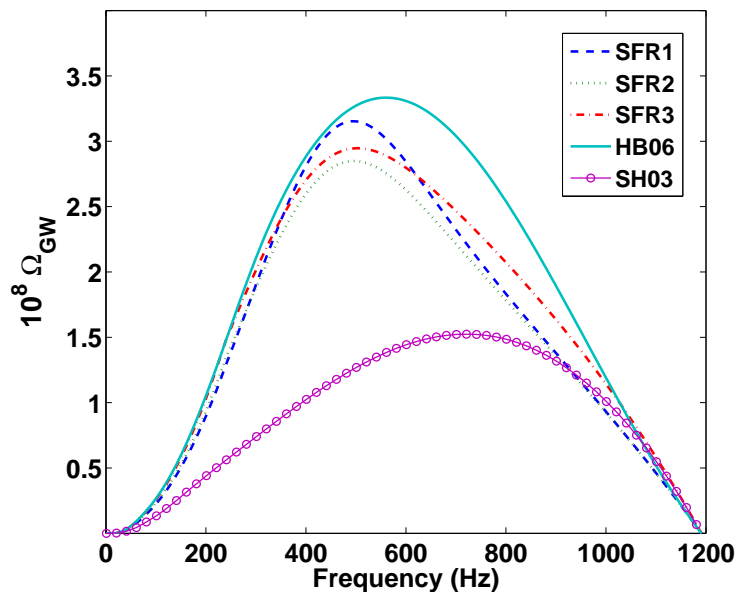} \figcaption{The dimensionless energy density
$\Omega_{\rm{GW}}$ as a function of observed frequency
$\nu_{\rm{obs}}$, calculated for five CSFR models and by setting $K
=-5/4$, $\nu_{\rm{max}}=1191$ Hz.}
\end{figure}

\begin{figure}
\plotone{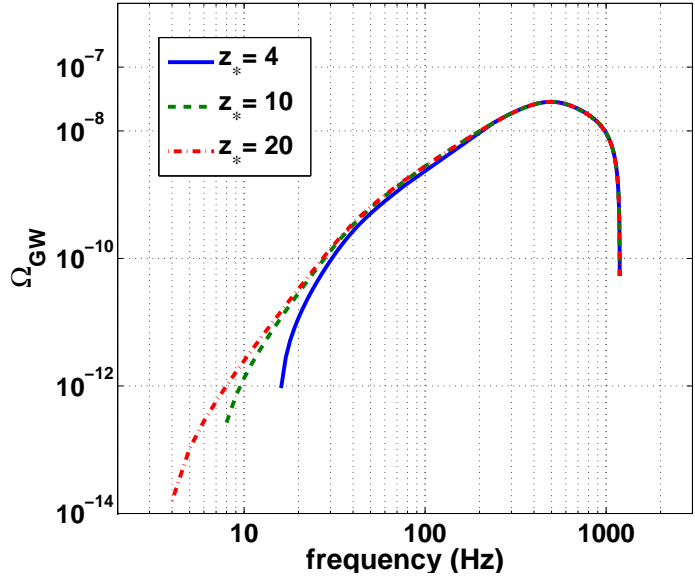} \figcaption{$\Omega_{\rm{GW}}$ as a function
of observed frequency $\nu_{\rm{obs}}$, calculated for SFR2 model
with three different values of maximum redshift $z_{\ast}$ and for
other parameters same as Fig. 3.}
\end{figure}

\begin{figure}
\plotone{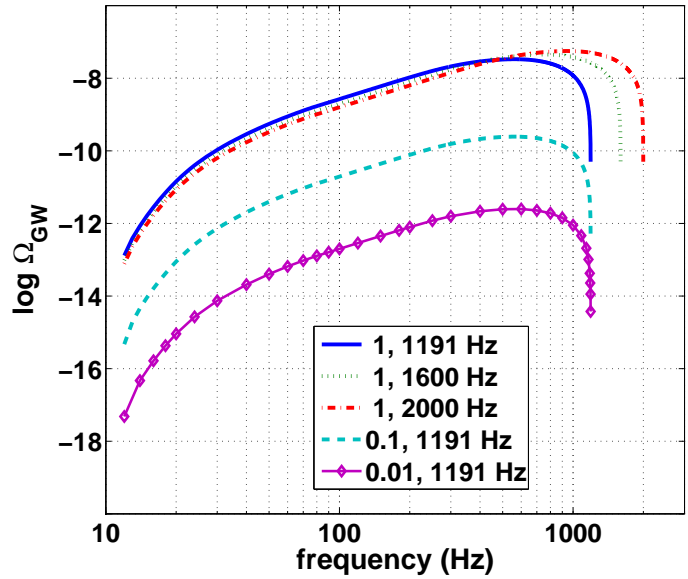} \figcaption{$\Omega_{\rm{GW}}$ as a function
of $\nu_{\rm{obs}}$, calculated for different values of ($\alpha$,
$\nu_{\rm{obs}}$).}
\end{figure}

\begin{figure}
\plotone{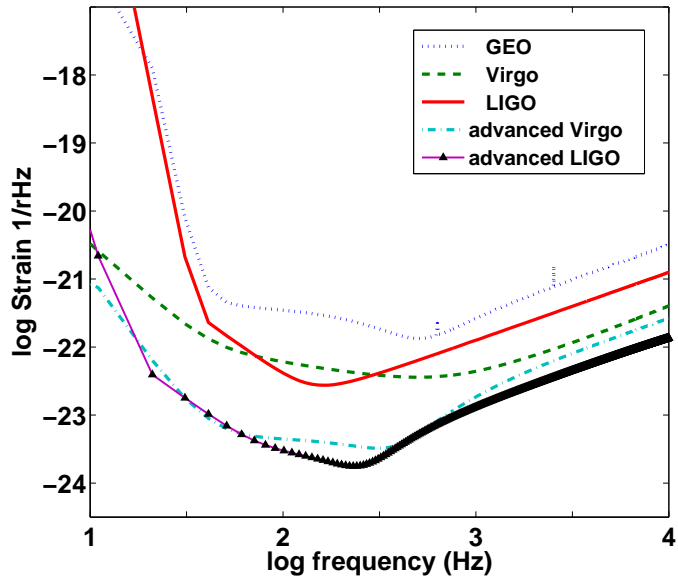} \figcaption{The designed noise power spectrum
of the GEO, Virgo, initial LIGO, advanced Virgo and advanced LIGO.
Data are taken from Lazzarini et al. 1996, Punturo 2004,
http://www.aei.mpg.de/jrsmith/geocurves.html,
http://wwwcascina.virgo.infn.it/advirgo/ and
https://dcc.ligo.org/cgi-bin/DocDB/ShowDocument?docid=2974.}
\end{figure}

\begin{figure}
\plotone{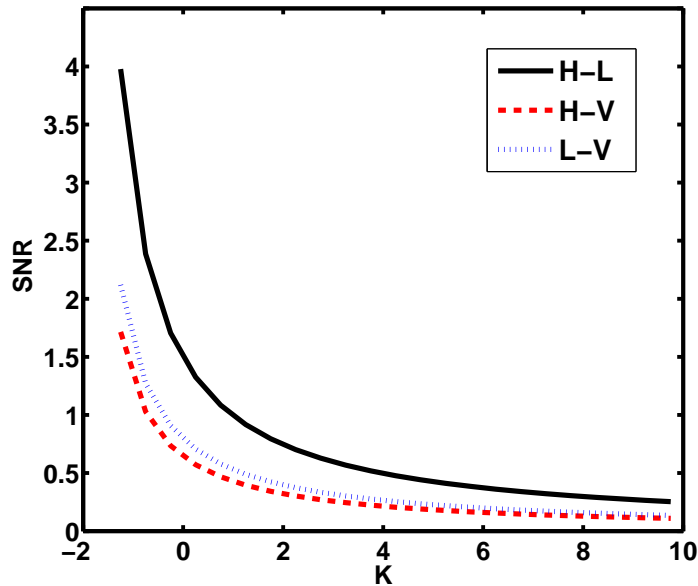} \figcaption{SNR as a function of $K$ for H-L,
H-V and L-V pairs when we set $\nu_{\rm{max}}=1191$ Hz. We assume a
factor of 10 folds improvement in sensitivity for advanced LIGO and
advanced Virgo detectors.}
\end{figure}

\begin{figure}
\plotone{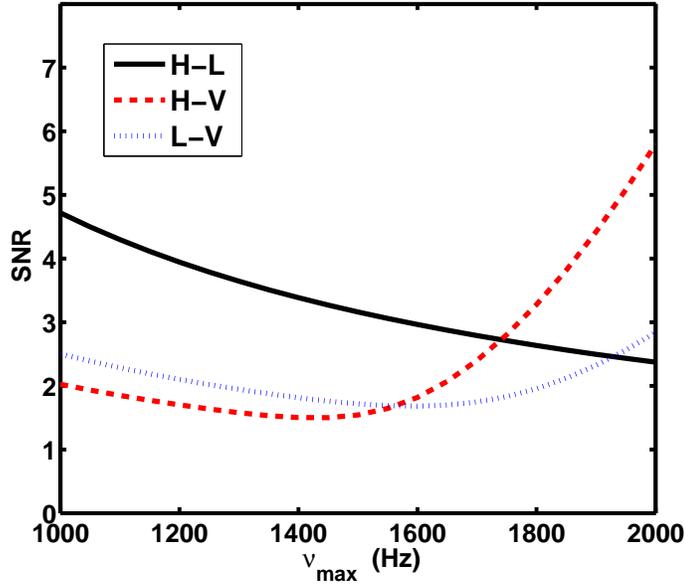} \figcaption{Same as Fig. 7, SNR as a function
of $\nu_{\rm{max}}$ for H-L, H-V and L-V pairs when we set
$K=-5/4$.}
\end{figure}

\begin{figure}
\plotone{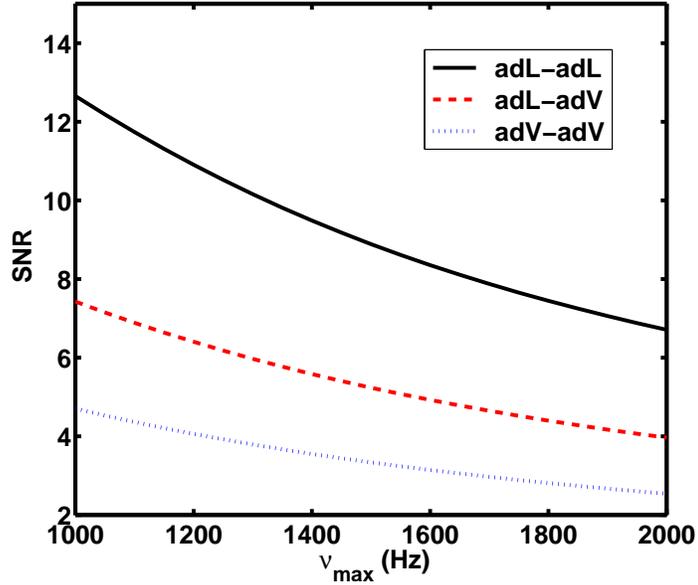} \figcaption{SNR as a function of
$\nu_{\rm{max}}$ assuming $\gamma(f)=1$ for two advanced LIGO (adL)
detectors, two advanced Virgo (adV) detectors and the combination
between them (here no improvement of sensitivity is assumed).}
\end{figure}

\begin{figure}
\plotone{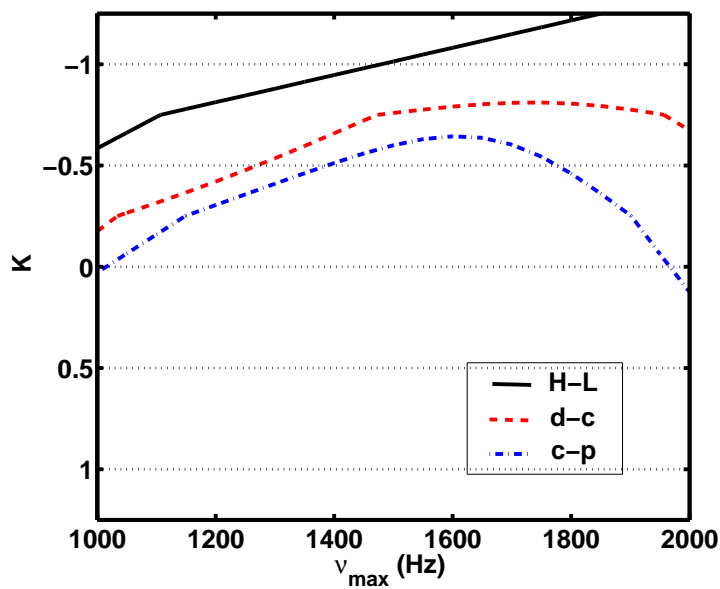} \figcaption{The region of parameter space for
which SGWB produced from an ensemble of NSs with r-mode instability
is detectable by H-L pair (here an order of magnitude improvement
for advanced LIGO sensitivity is assumed), the approach of directly
combining (labeled ``d-c") four third-generation detectors (an order
of magnitude improvement in detector sensitivities for Case 2 in
Table 2 is assumed) and the approach of combining multiple pairs of
detectors (labeled ``c-p"). With $10\%$ false alarm and $90\%$
detection rate, the region above the curves shows the detectable
parameter space.}
\end{figure}

\end{document}